\documentclass[aps,prb,amsmath,twocolumn]{revtex4-2}

\pdfoutput=1

\usepackage{amsmath}
\usepackage[mathlines]{lineno}%
\usepackage{dsfont}
\usepackage{mathtools}
\usepackage{caption}
\usepackage{graphicx}
\usepackage{epstopdf}
\usepackage{dcolumn}
\usepackage{bm}
\usepackage{color}
\usepackage{subfig}
\usepackage{float}
\usepackage[normalem]{ulem}
\usepackage{natbib}

\newcommand{\angstrom}{\textup{\AA}}
\newcommand{\subfigimg}[3][,]{%
	\setbox1=\hbox{\includegraphics[#1]{#3}}
	\leavevmode\rlap{\usebox1}
	\rlap{\hspace*{210pt}\raisebox{\dimexpr\ht1-2\baselineskip}{#2}}
	\phantom{\usebox1}
}

\newcommand{\modsubfigimg}[3][,]{%
	\setbox1=\hbox{\includegraphics[#1]{#3}}
	\leavevmode\rlap{\usebox1}
	\rlap{\hspace*{155pt}\raisebox{\dimexpr\ht1-2\baselineskip}{#2}}
	\phantom{\usebox1}
}

\providecommand{\abs}[1]{\lvert#1\rvert} 

\begin{document}

\title{Caroli formalism in near-field heat transfer \\ between parallel graphene sheets}

\author{Jia-Huei Jiang}
\email[Email: ]{u97810333@gmail.com}
\affiliation{Department of Physics, National Tsing-Hua University, Hsin-Chu 30013, Taiwan, ROC}
\author{Jian-Sheng Wang}
\affiliation{%
Department of Physics, National University of Singapore, Singapore 117551, Republic of Singapore
}%

\date{\today}


\begin{abstract}
In this work we conduct a close-up investigation into the nature of near-field heat transfer (NFHT) of two graphene sheets in parallel-plate geometry. We develop a fully microscopic and quantum approach using nonequilibrium Green's function method. A Caroli formula for heat flux is proposed and numerically verified. 
We show our near-field-to-black-body heat flux ratios generally exhibit $1/d^{\alpha}$ dependence, with an effective exponent $\alpha \approx 2.2$, at long distances exceeding 100 nm and up to one micron; in the opposite $d\rightarrow 0$ limit, the values converge to a range within an order of magnitude. We justify this feature by noting it is owing to the breakdown of local conductivity theory, which predicts a $1/d$ dependence. Furthermore, from the numerical result, we find in addition to thermal wavelength, $\lambda_{th}$, a shorter distance scale $\sim$ 10 - 100 nm, comparable to the graphene thermal length ($\hbar v_{F}/k_{B} T$) or Fermi wavelength ($k_{F}^{-1}$), marks the transition point between the short- and long-distance transfer behaviors; within that point, relatively large variation of heat flux in response to doping level becomes a typical characteristic. The emergence of such large variation is tied to relative NFHT contributions from the intra- and inter-band transitions.   Beyond that point, scaling of thermal flux $\propto 1/d^{\alpha}$ can be generally observed. 
\end{abstract}

\maketitle
\section{Introduction}

Within narrow vacuum gap compared with thermal wavelength  $\sim \lambda_{th}=\frac{\hbar\,c}{k_{B}T}$ between two bodies, surface modes can drastically augment electromagnetic thermal transfer by orders of magnitude greater than the normal Planckian radiative process\textemdash the so-called near-field heat transfer (NFHT).
The growing interest in NFHT ushers in novel designs of material systems and technological applications: thermal transistors \cite{Ben_Abdallah_2014}, thermal memory devices, thermophotovoltaic devices, thermal plasmonic interconnects \cite{Liu_2014}, and scanning thermal microscopy, just to name a few.  
Despite all the interesting designs of material properties and geometries, the description and starting point of the physical process has been centering around fluctuating current sources, about which the NFHT theory (the so-called ``fluctuational electrodynamics") was developed by Rytov \cite{Rytov}, and later formalized by Polder and Van Hove (PvH) \cite{Polder_1971}. Mahan has recently conducted similar inspection using two parallel metal surfaces \cite{Mahan_2017}. He compared the NFHT contributions from charge and current fluctuations and concluded that the former contribution is most important when the air gap between two surfaces is small. 

We are aware of some physical instances where the charge density fluctuation fits in more naturally than the current counterpart. Those instances are polar insulators, spatially confined nanostructures like nanodisks \cite{Yu_2017}, and graphene with plasmons \cite{Ilic_2012}. The third is the subject of this work. 

Charge density fluctuation due to thermal excitation and/or quantum effect gives rise to fluctuating electromagnetic fields. Because of 2D planar structure, graphene is credited for its great tunability of charge density and plasmonic excitability, which makes it an ideal material for close examination at the fluctuation of charge density. 

Owing to the fact that the plasmon wavelength $\lambda_{sp}$ is far shorter than thermal wavelength $\lambda_{th}=\frac{\hbar\,c}{k_{B}T}$, we can neglect retardation and attribute optical source fully in terms of the scalar potential \cite{Garc_a_de_Abajo_2014} $\phi$, which acts as the immediate field that couples to the charge density degrees of freedom. Herein with the exploration of NFHT to the nanometer and subnanometer scale \cite{sub_10nm_paper1, sub_10nm_paper2, sub_10nm_paper3, sub_10nm_paper4, sub_10nm_paper5}, the fully quantum description is needed. With the nonequilibrium nature of the transfer process, NEGF is versatile in coping with the scheme. Owing to the co-contribution made by Caroli, Combescot, Nozieres and Saint-James (CCNS), the so-called Caroli formula has become a handy tool in coping with the ballistic transport problem \cite{Caroli_1}. Despite of the handiness, its explicit use in previous works on NFHT has not been seen. In this work, we recognize the ballisticity of the NFHT process, under local equilibrium approximation (LEA) assumed here, and particularly, show a Caroli formula (see Eq.~(\ref{Caroli formula}) and Eq.~(\ref{Spectral Transmission}) in Sec.~\ref{sec: Method}). Since the trace for evaluating transmission is independent of geometry, though the parallel-plate geometry is given here as an example, the formula can be well applied to other geometries, e.g., tip-plane one, typical of scanning tunneling microscope. 

For visualization of field tunneling from one body through another, $\lambda_{th}=\hbar c/k_{B}T$ is a good figure of merit, smaller than which the near-field contribution dominates. However for graphene, a shrinkage of the characteristic length is notable because \cite{Vafek_2006} $v_{F}\sim c/300$. Moreover, other than the controlling factor $k_{B}T$, in cases where graphene is doped (meaning it has a finite chemical potential), the control factor shifts over to doping level (i.e. $k_{B}T \rightarrow \mu$ in the denominator of $\lambda_{th}$). In the following we will show a characteristic distance $\sim 10- 100$ nm, being comparable to $\hbar v_{F}/k_{B}T$ or $\hbar v_{F}/\mu$, and within and beyond which the different thermal flux behaviors show up.  

\section{Method \label{sec: Method}}

\begin{figure}[h]	
	\includegraphics[scale=0.2]{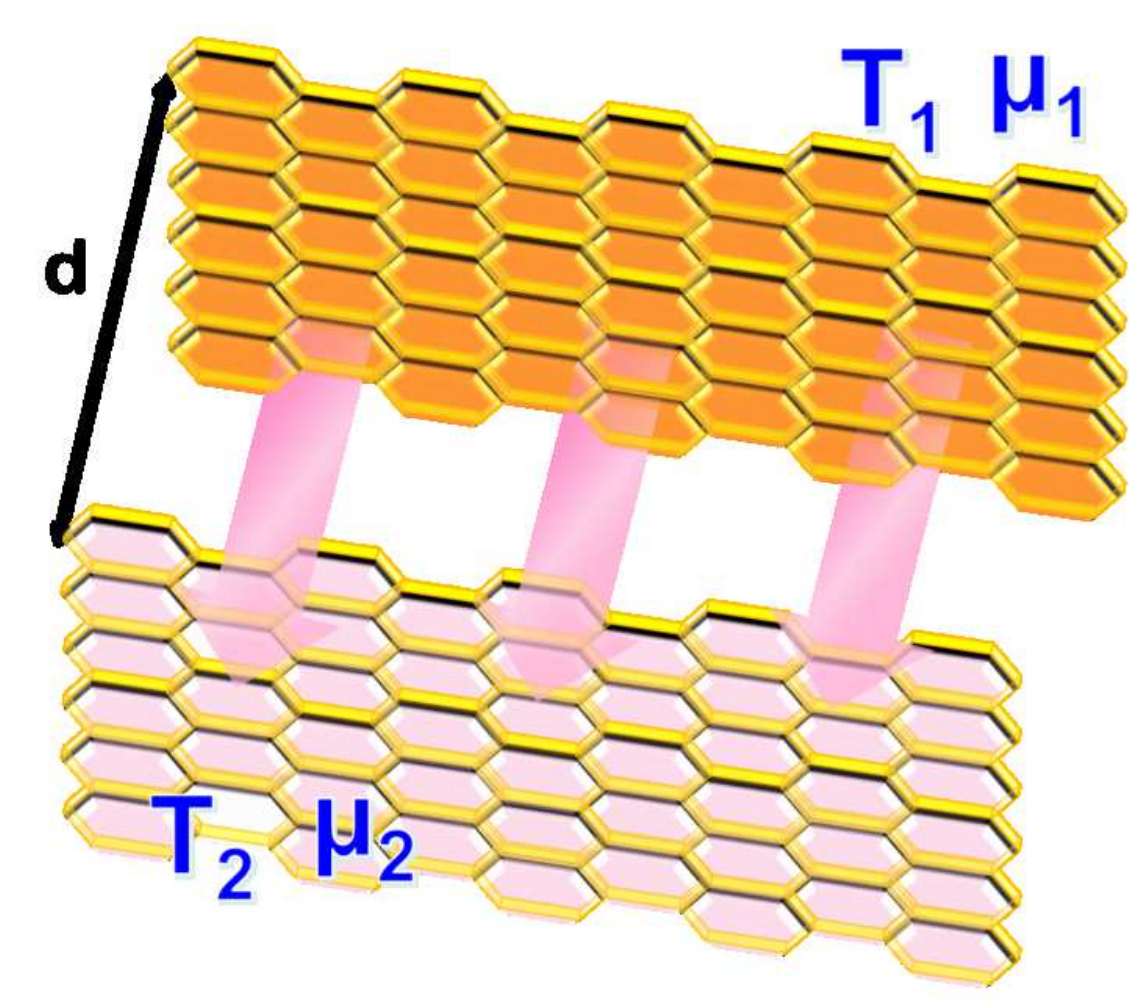}
	\captionsetup{justification=raggedright,
		singlelinecheck=false
	}
	\caption[]{\label{Fig_Graphene_NFHT}(Color online) The schematic showing heat transfer across vacuum gap of distance $d$ between two graphene sheets.}
\end{figure}

Consider two closely spaced graphene sheets with sheet 1 having temperature $T_{1}$, chemical potential $\mu_{1}$ and lying at $z=0$ plane and sheet 2 having temperature $T_{2}$, chemical potential $\mu_{2}$ and lying at $z=d$ plane (see Fig. \ref{Fig_Graphene_NFHT}). Starting from the scalar potential heat flux operator \cite{Peng_2-dot, Jiebin_EPL},
\begin{equation}\label{heat flux operator}
\hat{j}=\vdots\frac{\epsilon_{0}}{2}\left[\dot{\hat{\phi}}\,\nabla\hat{\phi}+\nabla\hat{\phi}\,\dot{\hat{\phi}}\right]\vdots
\end{equation}
with $\hat{\phi}$ denoting the field operator of scalar potential, $\epsilon_{0}$ the vacuum permittivity, sandwiching vertical dots the antinormal ordering, and dot above an operator the time derivative of that operator. The following NFHT formula in terms of the Green's function of scalar potential can be derived and reads
\begin{equation}\label{the proposed current flux}
\langle \hat{j}_z\rangle(z)=\, \epsilon_{0}\frac{1}{N}\sum_{ \textbf{q}_{\perp}} \int_{0}^{\infty}\frac{d\omega}{\pi}\,\hbar\omega\,
{\rm Re}\,\frac{\partial D_{jj}^{>}(\textbf{q}_{\perp},\omega,z,z')}{\partial \, z'}\biggr|_{z'=z},
\end{equation}
where $N$ is the total number of unit cells; $\textbf{q}_{\perp}=\left(q_{x},q_{y}\right)$ the 2D wavevector; $D_{j\,j'}^{>}(\textbf{q}_{\perp},\omega,z,z')$ is the Fourier transformed
\begin{equation}
\begin{aligned}
D_{j\,j'}^{>}&(\textbf{R},t,0,z,z')=\\
&-\frac{i}{\hbar}
\left<
\hat{\phi}_{j}(\textbf{R},z,t)
\hat{\phi}_{j'}(\textbf{0},z',0)\right>_{\mathcal{H}},
\end{aligned}
\end{equation}
the greater Green's function of scalar potential in $(\textbf{q}_{\perp},\omega)$ space; the ensemble average is taken with respect to the full Hamiltonian ${\mathcal{H}}$ defined in Appendix \ref{Hamiltonian}.
The formula is independent of $j=$ A or B, due to A-B sublattice symmetry.

Assuming local equilibrium for each of the sheets, the net heat flux, $J_{z}$, has a Caroli form: 

\begin{equation}\label{Caroli formula}
J_{z}={} \int_{0}^{\infty}\frac{d\omega}{2\pi}\hbar\omega
\left(N_{1}-N_{2}\right) 
T\left(\omega\right)
\end{equation}
with $N_{l}=1/(e^{\hbar\omega/k_{B}T_{l}}-1)$ being the Bose distribution at temperature $T_{l}$, and spectral transmission being 
\begin{equation}\label{Spectral Transmission}
T(\omega)=\int\frac{d^{2}\textbf{q}_{\perp}}{(2\pi)^2}\,
{\rm Tr}\left\{\hat{D}^{r}\hat{\Gamma}_{1}\hat{D}^{a}\hat{\Gamma}_{2}\right\},
\end{equation}
where we symbolically defined $\hat{O}$ as $4\times4$ plate- and sublattice-indexed matrices diagonal in $\left(\textbf{q}_{\perp},\omega\right)$ space, i.e., $\hat{O}=\hat{O}\left(\textbf{q}_{\perp},\omega\right)$. The trace is taken over plate and sublattice indices.

$\hat{D}^{r}$ is obtained from the Dyson equation:
\begin{equation}
\hat{D}^{r}=\hat{D}_{0}^{r}+\hat{D}_{0}^{r}\hat{\Pi}^{r}\hat{D}^{r}.
\end{equation}

The bare retarded Green's function for $\phi$, $D_{0}^{r}$, is
\begin{equation}\label{D0r}
D_{0}^{r}(\textbf{q}_{\perp},\omega,z,z')=
\frac{i\,e^{i\,q_{z}\,|z-z'|}}{2\,\epsilon_{0}\,S_{c}\,q_{z}}
\biggl(
\begin{matrix}
1 & e^{i\,\varphi}\\
e^{-i\,\varphi}& 1
\end{matrix}
\biggr),
\end{equation}
with
$q_{z}=\sqrt{\left(i\,\eta_{2}\right)^{2}-Q_{\perp}^{2}}$,
\,$\eta_{2}$ the damping term,  \,$Q_{\perp}=\sqrt{\frac{4}{3a_{0}^{2}}\left(3-|f(\textbf{q}_{\perp})|\right)}$,
\,$f(\textbf{q}_{\perp})=e^{-i\,q_{x}a_{0}}+e^{i\,q_{x}a_{0}/2+i\,\sqrt{3}q_{y}a_{0}/2}+e^{i\,q_{x}a_{0}/2-i\,\sqrt{3}q_{y}a_{0}/2}$,
\,$a_{0}$ the carbon-carbon distance,\, $S_{c}$ the unit-cell area,
\,$\varphi = -i\ln(\frac{f(\textbf{q}_{\perp})}{|f(\textbf{q}_{\perp})|})$.
$\hat{\Gamma}_{l}=i\,\left(\hat{\Pi_{l}}^{r}-\hat{\Pi_{l}}^{a}\right)$.

The appearance of $f$, $Q_{\perp}$ and $\varphi$ is the consequence of our approach of discretization of scalar potential on the graphene sheets (see Appendix \ref{Derivation of Retarded Bare Green's Function of Scalar Potential}). 

The Caroli formula, Eq.~(\ref{Caroli formula}), can be derived in several ways: one is to look at the work done by electric field on the current of the receiving sheet, i.e., joule heating, as analyzed by Yu et al. \cite{Yu_2017}.
Alternatively, one can equate field energy flowing into an enclosing surface with joule heating in its volume. Another is to consider Meir-Wingreen formula \cite{Meir-Wingreen_flormula} for the electron energy transfer \cite{ZZ_priv_comm}. Indeed, one can show our Caroli formula can be transformed exactly into Yu et al.'s form. Our formula is also consistent with Ilic et al.'s expression \cite{Ilic_2012} in the non-retarded limit.

\begin{figure}
	\includegraphics[width=\columnwidth]{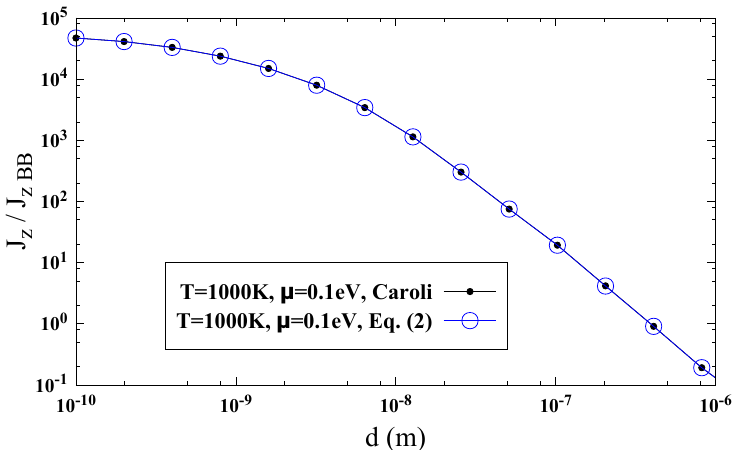}
	\captionsetup{justification=raggedright,
		singlelinecheck=false
	}
	\caption{\label{Proof of Caroli formula}(Color online) A comparison of the results calculated with Caroli formula and Eq.~(2). The close match is a proof of equivalence.}
\end{figure}

The equivalence between the Caroli formula [Eq.~(\ref{Caroli formula})] and Eq.~(\ref{the proposed current flux}) can be numerically checked. We see perfect match in Fig.~\ref{Proof of Caroli formula}. In addition, an analytical proof of the equivalence in long-wave limit is provided in Appendix \ref{Analytic Proof of the Caroli Formula in Long Wave Limit}. 

The self-energy $\Pi^{r}$ is evaluated in the random phase approximation (RPA) \cite{Wunsch_2006,Hwang_2007} and local equilibrium approximation (LEA). The form is given in Eq.~(\ref{Self_retarded}).

Last, some words on numerical implementation. When evaluating Eq.~(\ref{Self_retarded}), one might have thought of the possibility that convolution integrals in $\left(\textbf{k},\nu\right)$ space can be avoided, by a Fast Fourier Transform (FFT) of $\Pi^{r}$ from $\left(\textbf{R},t\right)$ space to $\left(\textbf{q}_{\perp},\omega\right)$ space. Experience told us that using FFT, though fast when giving $\Pi^{r}$, becomes computationally resource demanding as one tries to cover the long-wave contribution ($\textbf{q}_{\perp} \rightarrow 0$) of field.
When we calculate Eq.~(\ref{Spectral Transmission}), isotropy (i.e., the integrand depends only on $|\textbf{q}_{\perp}|$) is assumed for $\textbf{q}_{\perp}$ integral. Even though we discretize scalar potential on the lattice (Appendix~\ref{Derivation of Retarded Bare Green's Function of Scalar Potential}), only when in high-energy region would anisotropy of field dispersion become significant.  We have compared the results calculated using such approximation with the original ones using no approximation and found no essential difference.
Also, to save computational time, suitable cutoffs for integrals can be employed. Whether a cutoff is explicitly taken or not does not affect physics because the natural limit set by Boltzmann factor. Great care has been taken in regard to this part to make sure there is no loss of key information.

\section{Results and discussion}
\begin{figure}[h]
	\includegraphics[width=\columnwidth]{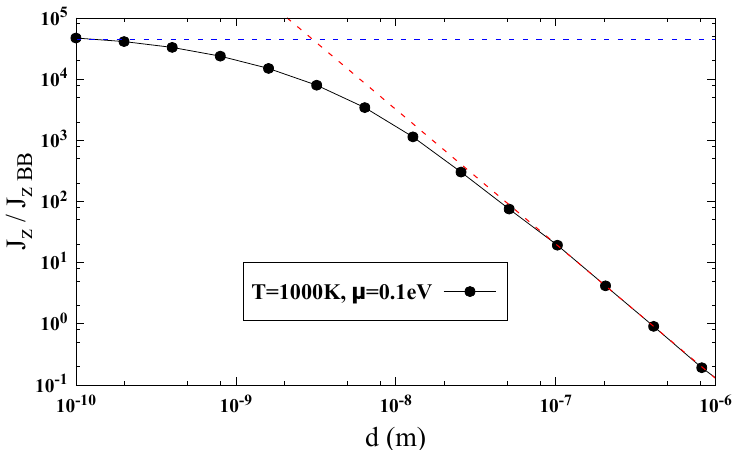}
	\captionsetup{justification=raggedright,
		singlelinecheck=false
	}
	\caption[]{\label{2_asymptotes}(Color online) The red dashed line indicates the asymptotic behavior $1/d^{\alpha}$, $\alpha \approx 2.2$ as $d$ becomes of micrometer scale. The blue dashed line indicates the saturation of the curve when $d$ approaches zero. $J_{z BB} = 56244$ W\,$m^{-2}$.}
\end{figure}

Figure~\ref{2_asymptotes} demonstrates the calculated heat flux ratio over the blackbody limit. Plate 1 has temperature $T_{1}$ = 1000 K,chemical potential $\mu_{1}$ = 0.1 eV; plate 2 has temperature $T_{2}$ = 300 K, chemical potential $\mu_{2}$ = 0.1 eV. We set damping factor of electron $\eta_{1} = 0.0033$ eV (corresponding to the life time of electron $\tau \sim 10^{-13}$ sec) for both sheet 1 and 2 throughout this work. The red dashed line corresponds to the $1/d^{2.2}$ scaling of the flux $J_{z}$ when $d$ is large (beyond a few ten nanometers). Without further analytic support for the exponent $2.2$, we can only view it as an effective exponent $\alpha$. Some might expect $\alpha$ be some simpler value like $2$, which has been predicted by Loomis and Maris \cite{one_over_d_squared}, when discussing dielectrics with pointlike dielectric constants. The dielectric constant of graphene sheet is no way pointlike, as due to its plasmon activity. Thus some correction (e.g. $0.2$ as found here) to $2$ is expected. The left end of the blue dashed line points to a convergent value at $J_{z}/J_{z BB}$ $\approx 46660$. Though at/below $d \approx a_{cc} \sim 1 \angstrom$ it might reach the contact limit, our calculations at/below that distance indicate asymptotically convergent values.

\begin{figure}[h]	
    \subfloat{\subfigimg[width=\columnwidth]{\textbf{(a)}}{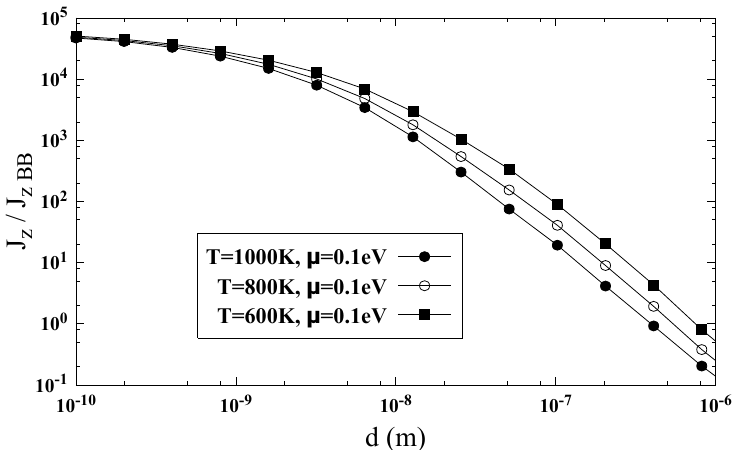}\label{temp_var}}\\
	\subfloat{\subfigimg[width=\columnwidth]{\textbf{(b)}}{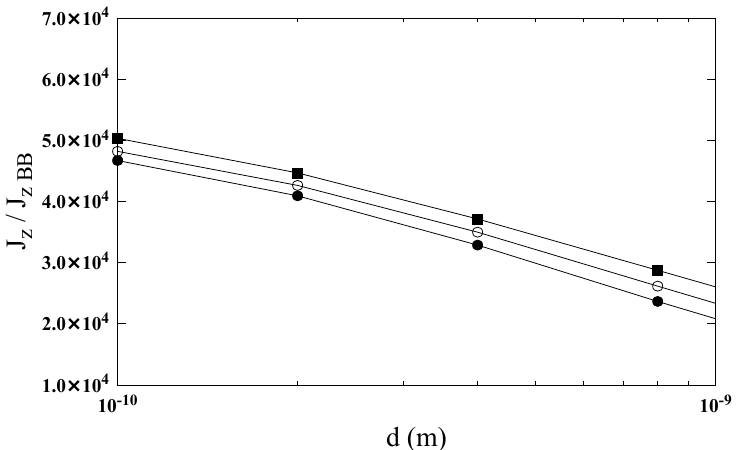}\label{temp_var_zoom_in}}
	\captionsetup{justification=raggedright,
		singlelinecheck=false
	}
	\caption[]{(a) Heat flux ratio in different temperatures ($\mu = 0.1$ eV). (b) The short-distance zoom-in.}
\end{figure}

In Fig.~\ref{temp_var} we set various temperatures (600, 800, and 1000 K) on sheet 1 with the temperature on sheet 2 kept at 300K; the chemical potential on both sheets is 0.1 eV. The flux ratio generally decreases with increasing temperature at a given distance because the fourth power of $T$ in the denominator of flux ratio.

Notably, the results recover the two features mentioned: the $d\rightarrow 0$ asymptotic convergence values $\sim 5\times 10^{4}$ and $1/d^{\alpha}$ scaling at long distances.   

\begin{figure}[ht]
	\includegraphics[width=\columnwidth]{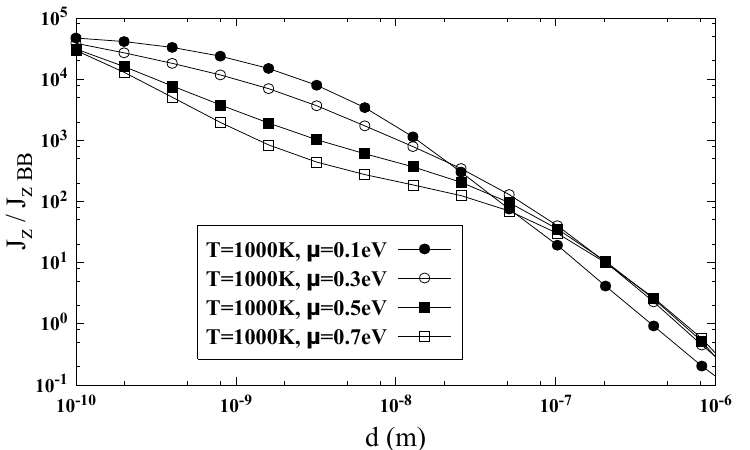}
	\captionsetup{justification=raggedright,
		singlelinecheck=false
	}
	\caption[]{\label{chemipo_var}The heat flux in different chemical potentials.}
\end{figure}

We then compare the effect of doping on flux ratios. By doping level (or chemical potential) we mean the same doping level on both sheets; identical surfaces except for temperature difference were shown to achieve maximal NFHT \cite{Ilic_2012}. 
In the $d\rightarrow 0$ limit, we note, interestingly, the heat flux ratios of all doping levels converge to values $\sim 5\times 10^{4}$.
It can also be noted that roughly below a few ten nanometers the flux ratio with higher doping levels (e.g. 0.7 eV) exhibit a lower-lying arch of flux ratios. This is because the interband transition gap is opened by doping. For light doping level such as 0.1 eV, it is easier for interband transition to occur \cite{Wunsch_2006,Hwang_2007}, thus the straddling upper arch (Fig.~\ref{chemipo_var}). On the other hand, beyond 100 nm, the large modulation in heat flux in response to doping level is no longer seen\textemdash the curves become a constricted stream having a scaling $\propto 1/d^{\alpha}$. We nickname the typical shape formed by the straddling and lower-lying arches the ``doping bubble".

The distance at a few ten nanometers, separating the doping bubble at short distances, and the $1/d^{\alpha}$ stream at long distances, is reminiscent of the Vafek's thermal length of graphene ($\hbar v_{F}/k_{B}T$) in the high temperature limit, or Fermi wavelength ($k_{F}^{-1}$) in the high doping level limit \cite{Vafek_2006,G_mez_Santos_2009,Svetovoy_2011}. It is tempting to associate the distance with either $\hbar v_{F}/k_{B}T$ or $k_{F}^{-1}$, for within the parameter set chosen the two length scalings both give an estimate $\sim$ 10 - 100 nm (e.g. $\mu_{1} \approx 1 \,T_{1}, \,3 T_{1}, 5 \,T_{1}, 7 \,T_{1}$ when $T_{1}$ = 1000 K; $\mu_{2} \approx 3 \,T_{2}, \,9 T_{2}, 15 \,T_{2}, 35 \,T_{2}$ when $T_{2}$ = 300 K). However, the situation is more complex than that, for the finite temperature difference across two sheets and accordingly, different weights on temperature and doping level for each sheet. As such, the distance may be deemed just as order-of-magnitude estimate in NFHT problem.

Doping bubble holds the possibility for dynamic control of NFHT below thermal wavelength.

\begin{figure}[h]	
	\includegraphics[width=\columnwidth]{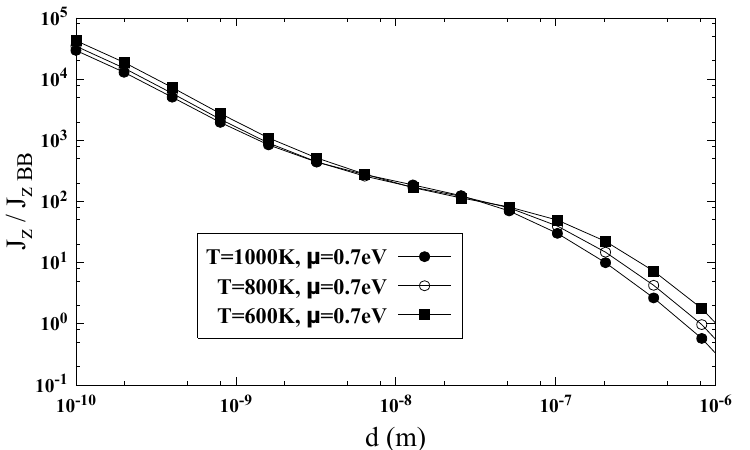}
	\captionsetup{justification=raggedright,
		singlelinecheck=false
	}
	\caption[]{\label{temp_var2}Heat flux ratio in different temperatures ($\mu = 0.7$ eV).}
\end{figure}

We further examine the doping bubble and the characteristic distance in the higher doping case (0.7 eV, as selected) across different temperatures. The result is presented in Figs.~\ref{temp_var} and \ref{temp_var2}. They show the curves are subject only to minor changes when temperature is varied. The change is smaller in Fig.~\ref{temp_var2} than in Fig.~\ref{temp_var} because of larger weight on doping level controlling graphene plasmonic excitation.  

As can also be seen in Figs.~\ref{temp_var_zoom_in}, \ref{chemipo_var}, and \ref{temp_var2}, these two asymptotes are insensitive to temperature and chemical potential variation over the parameters chosen, implying they pose as a general asymptotic feature in NFHT using two-graphene-plate geometry.  

\begin{figure}[h]	
	\includegraphics[width=\columnwidth]{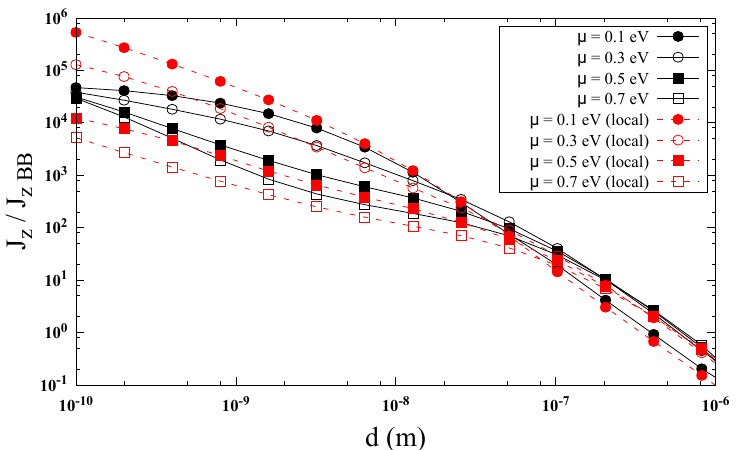}
	\captionsetup{justification=raggedright,
		singlelinecheck=false
	}
	\caption[]{\label{Nonlocal&local_Flux}(Color online) Comparison of heat flux calculated using the full RPA and local conductivity models.}
\end{figure}

Last, we use Ilic et al.'s method with spatially local conductivity \cite{Ilic_2012} and plot curves to compare with what were already shown in Fig.~\ref{chemipo_var}. Such comparison is shown in Fig.~\ref{Nonlocal&local_Flux}.

\begin{figure}[hb]
	\subfloat{\modsubfigimg[width=0.95\columnwidth,height=0.73\columnwidth]{\textbf{\textcolor{white}{(a)}}}{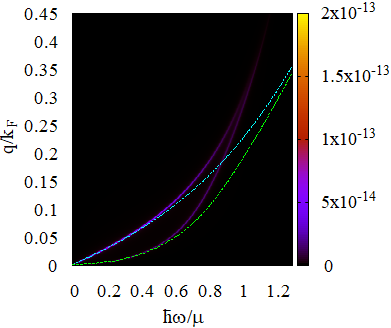}\label{Disp_mine_0.7_10nm}}\\
	\subfloat{\modsubfigimg[width=0.95\columnwidth,height=0.73\columnwidth]{\textbf{\textcolor{white}{(b)}}}{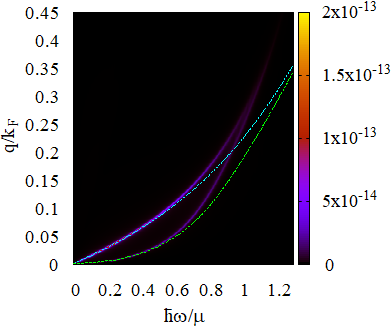}\label{Disp_Ilic_0.7_10nm}}
	\captionsetup{justification=raggedright,
		singlelinecheck=false
	}
	\caption[]{\label{Nonlocal&local_Disp_10nm}(Color online) Match in plasmon dispersion between the full RPA and local conductivity models at $d = 10$ nm ($\mu =$ 0.7 eV). Transmission $T(\omega)$ plots are weighted by $q\,\hbar\,\omega\,(N_{1}-N_{2})$. (a) The full RPA calculation. (b) Local conductivity calculation. Cyan and green dashed lines stand for acoustic and optical modes in $q \rightarrow 0$ limit, respectively.}
\end{figure}

\begin{figure}[ht]
	\subfloat{\modsubfigimg[width=0.95\columnwidth,height=0.73\columnwidth]{\textbf{\textcolor{white}{(d)}}}{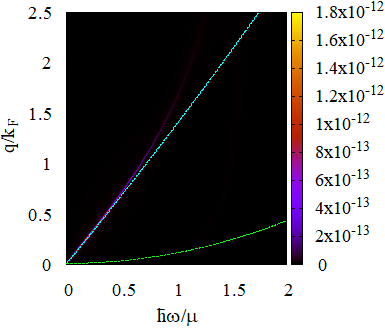}\label{Disp_Ilic_0.7}}
	\captionsetup{justification=raggedright,
		singlelinecheck=false
	}

	\caption[]{\label{Nonlocal&local_Disp}(Color online) Disparity in plasmon dispersion between the full RPA and local conductivity models at $d = 1 \angstrom$.  Transmission $T(\omega)$ plots are weighted by $q\,\hbar\,\omega\,(N_{1}-N_{2})$. (a) and (c): The full RPA calculations with $\mu =$ 0.1 and 0.7 eV, respectively. (b) and (d): Local conductivity calculations with $\mu =$ 0.1 and 0.7 eV, respectively. Cyan and green dashed lines stand for acoustic and optical modes in $q \rightarrow 0$ limit, respectively.}
\end{figure}

	As one can see,  when $d \rightarrow 0$ the clear disparity between our full RPA and local conductivity calculations shows up, but as $d$ goes larger the disparity gradually vanishes. The local conductivity model does show $1/d$ heat flux dependence as $d \rightarrow 0$\, \cite{0953-8984-27-21-214019,PhysRevB.92.144307}. In contrast, our calculation shows saturated value for $\mu =$ 0.1 eV and ``augmented" value for $\mu = 0.7$ eV. To explain the disparity at extremely small $d$'s, we examine the plots of transmission $T(\omega)$ weighted by $q\,\hbar\,\omega\,(N_{1}-N_{2})$ in Fig.~\ref{Nonlocal&local_Disp_10nm} and \ref{Nonlocal&local_Disp}. First from Fig.~\ref{Nonlocal&local_Disp_10nm}, we see close match between the full RPA and local conductivity calculations at $d = 10$ nm. Such fact shows the validity of the full RPA calculations in large-distance limit.

\begin{figure}[h]
	\subfloat{\modsubfigimg[width=0.95\columnwidth,height=0.73\columnwidth]{\textbf{\textcolor{white}{(a)}}}{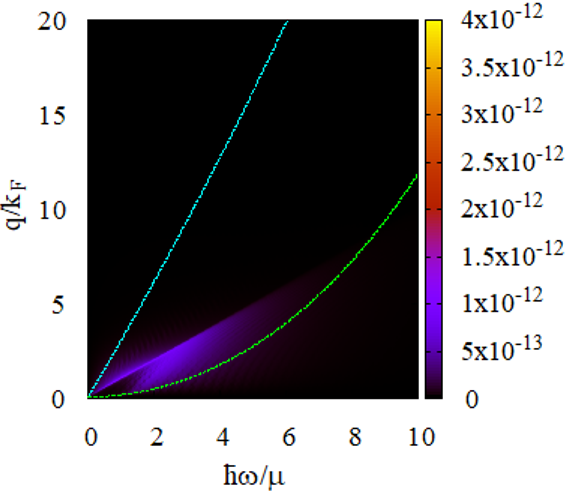}\label{Disp_mine_0.1}}\\
	\subfloat{\modsubfigimg[width=0.95\columnwidth,height=0.73\columnwidth]{\textbf{\textcolor{white}{(b)}}}{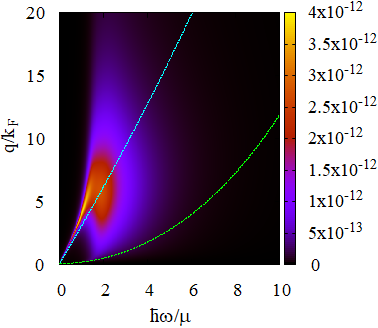}\label{Disp_Ilic_0.1}}\\
	\subfloat{\modsubfigimg[width=0.95\columnwidth,height=0.73\columnwidth]{\textbf{\textcolor{white}{(c)}}}{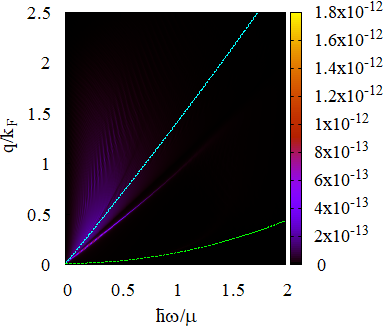}\label{Disp_mine_0.7}}

	\caption*{}
\end{figure}

	Second, from Fig.~\ref{Nonlocal&local_Disp}, we find a mismatch in plasmon dispersion at $d = 1 \angstrom$. In particular, the greatest contributing branches \cite{PhysRevB.92.144307}\textemdash acoustic plasmon modes differ in slope in $\omega-q$ plane, even in long-wave limit (the arches of optical modes, which are expected $d$-independent in long-wave limit in local conductivity theory, are nearly the same in the two cases (not clearly shown due to range of colorbox chosen)). In spite of the mismatch, we can also note that acoustic branch of local conductivity model extends into the region: $\omega < v_{F}q$ that meets the failure for best description by local conductivity ($\omega > v_{F}q$) \cite{Falkovsky_1,Falkovsky_2}. Such breakdown of local conductivity theory is evident because acoustic mode frequency $\omega_{L} \propto \sqrt{d}$. When $d$ gets small enough, that frequency no longer satisfies $\omega_{L} > v_{F}q$. For $\mu =$ 0.1 eV, the critical distance $d_{c} \simeq 1.24$ nm; for $\mu =$ 0.7 eV, $d_{c}\simeq 1.77 \angstrom$. The full RPA calculation of ours has rescued the extinction of acoustic plasmon mode under local conductivity approximation by constraining the mode to stay within border of $\omega = v_{F}q$ line (see Figs.~\ref{Disp_mine_0.1} and \ref{Disp_mine_0.7}). Comparatively, Fig.~\ref{Disp_mine_0.1} has apparent small span in $q$ and less spectral weight. This accounts for the saturation. And judging from Figs.~\ref{Disp_mine_0.7} and 9d , though the $\omega_{L}$ line of the full RPA appears a little bit shorter than that of local conductivity's, it has a wide fan-out area in $\omega < v_{F}q$, thus the augmented value in $J_{z}$ comparatively.


\section{Conclusion}
In this work, an NEGF based theory was proposed to analyze NFHT. For the ease of analysis we take the first step to consider the scalar-potential-mediated NFHT of graphene in parallel-plate geometry. A Caroli formula for the heat transfer was derived from a scalar-potential-based heat flux inspired by our previous works on NFHT.

The density-density correlation (self-energy of scalar potential) was derived within RPA without taking long-wave approximation. By following LEA, we create a platform to compare our approach with the former works that succeeded from Rytov's theory.

We found, numerically, three notable features: (i) in $d\rightarrow 0$ limit flux ratio curves with all doping levels and temperature range selected converge to a limited range of values  $\sim 10^{4}- 10^{5}$ within an order of magnitude variation. (ii) the existence of a highly doping-tunable region dubbed ``doping bubble" lying between that limit and $d \approx 100$ nm. (iii) Beyond 100 nm the curves all possess a $1/d^{\alpha}$ scaling ($\alpha \approx 2.2$) and the large variation of flux ratio as within 100 nm vanishes. (i) and (ii) stand for the $d \rightarrow 0$ behavior that local conductivity theory cannot capture. In regard to (iii), the range 100 nm is close to and reminiscent of Vafek's ``thermal length" and Fermi wavelength \cite{Vafek_2006,G_mez_Santos_2009,Svetovoy_2011}. 
However due to finite temperature difference and different weights on temperature as well as doping level, direct attribution is invalid. Thus we deem it as an estimate scale correct within an order of magnitude.

Since our NEGF approach marks the full quantum mechanical feature, with the ever-deepened exploration down to $< 10$ nm, characteristic result like doping bubble sitting within small plate-plate distance range $\sim 1 \angstrom - 100$ nm that we found in this work can be directly tested. The graphene's large modulation of flux with different doping levels down to nanoscale holds possibility for future active nanothermal management. 

\section*{Acknowledgment}

J.-S.W. was supported by FRC grant No. R-144-000-343-112.


\appendix

\section{Hamiltonian}\label{Hamiltonian}
The quantum Hamiltonian is given by
\begin{equation}
\begin{aligned}
&\mathcal{H}=\mathcal{H}_{\phi}+\mathcal{H}_{e}+\mathcal{H}_{int},\\
& \mathcal{H}_{\phi}= -\frac{\epsilon_{0}}{2}\int dV
\left[\left(\frac{\dot{\hat{\phi}}}{\tilde{c}}\right)^{2}+\left(\nabla\hat{\phi}\right)^{2}\right],\\
& \mathcal{H}_{e}=
\sum_{\textbf{k},\,l = 1,2} 
{c_{\textbf{k}}}^{\dagger\,(l)}\biggl[
\begin{matrix}
0 & -\gamma_{0}f(\textbf{k})\\
-\gamma_{0}f(\textbf{k})^{*} & 0
\end{matrix}
\biggl]c_{\textbf{k}}^{(l)},\\
& \mathcal{H}_{int}=
\sum_{\substack{\textbf{R},\;l = 1,2\\j=A,B}}
-e\,\hat{\phi}_{j}(\textbf{R},z^{(l)})\,{c_{j\,\textbf{R}}}^{\dagger\,(l)}c_{j\,\textbf{R}}^{(l)}, 
\end{aligned}
\end{equation}
where $\hat{\phi}$ denotes scalar potential field operator;
\,$c_{\textbf{k}}^{(l)}=\left(c_{A\,\textbf{k}}^{\quad (l)},\, c_{B\,\textbf{k}}^{\quad (l)}\right)^{T}$;
\,${c_{\textbf{k}}}^{\dagger\,(l)}=\left({c_{A\,\textbf{k}}}^{\dagger\,(l)},\, {c_{B\,\textbf{k}}}^{\dagger\,(l)}\right)$;
\,$\gamma_{0}$ = 2.8 eV;
\,$f(\textbf{k})=e^{-i\,k_{x}a_{0}} +\,e^{i\,k_{x}a_{0}/2+i\,\sqrt{3}k_{y}a_{0}/2}+
\,e^{i\,k_{x}a_{0}/2-i\,\sqrt{3}k_{y}a_{0}/2}$;
\,$\varphi(\textbf{k}) = -i \ln(f(\textbf{k})/|f(\textbf{k})|)$;
\,$l$ is the plate index.

The electronic Hamiltonian $\mathcal{H}_{e}$ is assumed by a tight-binding model. The Hamiltonian for vector potential does not enter because of the quasistatic limit. A real parameter $\tilde c$ is initially kept finite in the Hamiltonian of scalar potential $\mathcal{H}_{\phi}$ for the ease of quantization. After the bare scalar potential Green's function is evaluated, we can go back and continue on our quasistatic approximation by simply forsaking $\omega$ dependence in $q_{z}$ entirely, i.e.,
\begin{equation*}
\begin{aligned} 
q_{z}={}&\lim_{\tilde c\rightarrow \infty}\sqrt{\left(\frac{\omega}{\tilde c}+i\,\eta_{2}\right)^{2}-Q_{\perp}^{2}}=i\,\sqrt{\eta_{2}^{2}+Q_{\perp}^{2}}. 
\end{aligned}
\end{equation*}
\section{Derivation of Bare Retarded Green's Function of Scalar Potential}\label{Derivation of Retarded Bare Green's Function of Scalar Potential} 
We define the bare retarded scalar potential Green's function as
\begin{equation} \label{bare retarded scalar potential Green's function 1}
\begin{aligned}
D_{0\,j\,j'}^{r}&(\textbf{R},t,0,\,z,z')=\\
&-\frac{i}{\hbar}\theta(t)\left<\left[
\hat{\phi}_{j}(\textbf{R},z,t),\hat{\phi}_{j'}(\textbf{0},z',0)\right]\right>_{\mathcal{H}_{\phi}}
\end{aligned}
\end{equation}
where $\textbf{R}$ is the transverse lattice vector and $\left\{ j,j' \right\}=\left\{ A,B \right\}$; $\left[a,\,b\right]$ is the commutator of operator $a$ and $b$.
Equation~(\ref{bare retarded scalar potential Green's function 1}) can be easily derived by the equation of motion method. But prior to that, we discretize scalar potential on the graphene lattice in directions parallel to the sheets (the transverse directions) as an approximation to ease the calculation. The field in the direction perpendicular to the planes (the $z$ direction) is still treated as continuous. The approximation makes sense in that we consider only the field generated by the fluctuating density of electron on one sheet and transmitted energy is maximally absorbed by another sheet.
The equation of motion for the Green's function is then,
\begin{equation}
\begin{aligned}
&\sum_{i}\left[
\biggl(\frac{1}{{\tilde c}^{2}}\frac{\partial^{2}}{\partial t^{2}}
-\frac{\partial^{2}}{\partial z^{2}}\biggr)\delta_{ji}
-\left[\nabla_{\perp}^{2}\right]_{ji}(\textbf{R})
\right] \\ 
&D_{0\,i\,j'}^{r}(\textbf{R},t,0,\,z,z')=\\ &\frac{1}{\epsilon_{0}}\delta(t)\left(\frac{2}{S_{c}}\delta_{\textbf{R},\,\textbf{0}}\,\delta_{j\,j'}\right)\delta(z-z').
\end{aligned}
\end{equation}

The factor $2$ on the right hand side accounts for the subdivision of $A$ and $B$ sublattices.

The Laplacian operator $\sum_{i}\left[\nabla_{\perp}^{2}\right]_{ji}(\textbf{R})$ has to obey lattice periodicity and is defined by
\begin{equation}
\begin{aligned}
\left[\nabla_{\perp}^{2}\right](\textbf{R})\:
\hat{\phi}_{A}(\textbf{R})=\frac{4}{3\,a_{0}^{2}}
&\biggl[\hat{\phi}_{B}(\textbf{R})+\hat{\phi}_{B}(\textbf{R}+ \textbf{a}_{1})+\\
&\hat{\phi}_{B}(\textbf{R}+ \textbf{a}_{1} - \textbf{a}_{2})-3\,\hat{\phi}_{A}(\textbf{R})\biggr],\\
\left[\nabla_{\perp}^{2}\right](\textbf{R})\:
\hat{\phi}_{B}(\textbf{R})=\frac{4}{3\,a_{0}^{2}}
&\biggl[\hat{\phi}_{A}(\textbf{R})+\hat{\phi}_{A}(\textbf{R}- \textbf{a}_{1})+\\
&\hat{\phi}_{A}(\textbf{R}- \textbf{a}_{1} + \textbf{a}_{2})-3\,\hat{\phi}_{B}(\textbf{R})\biggr],
\end{aligned}
\end{equation}
or equivalently in $\textbf{q}_{\perp}$ space
\begin{equation}
\begin{aligned}
&\left[\nabla_{\perp}^{2}\right](\textbf{q}_{\perp})\,
\left[\begin{matrix}
\hat{\phi}_{A}(\textbf{q}_{\perp}) \\
\hat{\phi}_{B}(\textbf{q}_{\perp})
\end{matrix}\right]=\\
&\left[\frac{4}{3\,a_{0}^{2}}\right]
\left[\begin{matrix}
-3 && f(\textbf{q}_{\perp})\\
f^{*}(\textbf{q}_{\perp}) && -3
\end{matrix}\right] 
\left[\begin{matrix}
\hat{\phi}_{A}(\textbf{q}_{\perp}) \\
\hat{\phi}_{B}(\textbf{q}_{\perp})
\end{matrix}\right].
\end{aligned}
\end{equation}
$\textbf{a}_{1}=\left(\frac{3}{2},\,\frac{\sqrt{3}}{2}\right)a_{0}$, $\textbf{a}_{2}=\left(0,\,\sqrt{3}\right)a_{0}$.
It is easy to check the Laplacian operator so defined is valid in the long-wave approximation.

As such, the equation of motion in $(\textbf{q}_{\perp},\omega)$ space reads
\begin{equation}
\begin{aligned}
&\sum_{i}\left[
\left(\left(\frac{\omega}{\tilde c}+
i\,\eta_{2}\right)^{2}+
\frac{\partial^{2}}{\partial z^{2}}\right)\delta_{ji}
+\left[\nabla_{\perp}^{2}\right]_{ji}(\textbf{q}_{\perp})\right]\\ &D_{0\,i\,j'}^{r}(\textbf{q}_{\perp},\omega,\,z,z')=
-\left(\frac{2}{\epsilon_{0}\,S_{c}}\,\delta_{j\,j'}\right)\delta(z-z').
\end{aligned}
\end{equation}
$\eta_{2}$ is the damping factor of scalar potential. Such damping factor should be small; we set $\eta_{2} \sim 10^{-5} \, \mathrm{m}^{-1}$ . Taking the inverse of the operator matrix on the left hand side and following complex integration, we finally get Eq.~(\ref{D0r}) in the main text.
\section{Evaluation of Self-energy}
\subsection{RPA}

Consider only one of the two sheets, the RPA retarded self-energy, in $(\textbf{k}, \nu)$ space reads \cite{Peng_2-dot,Jiebin_EPL}
\begin{equation} \label{Self_retarded}
\begin{aligned}
\Pi^{r}_{jj'}(\textbf{q}_{\perp},\omega)= &
-\frac{2\,i\,\hbar\,e^{2}}{N}\sum_{\textbf{k}} \int\frac{d\nu}{2\pi}\\
&\biggl\{G_{jj'}^{r}(\textbf{k},\nu)\,G_{j'j}^{<}(\textbf{k}-\textbf{q}_{\perp},\nu-\omega)\\
+&\,G_{jj'}^{<}(\textbf{k},\nu)\,G_{j'j}^{a}(\textbf{k}-\textbf{q}_{\perp},\nu-\omega)\biggr\}.
\end{aligned}
\end{equation}
A prefactor of 2 accounts for spin degeneracy. $j,\,j'=A, B$.

Substitute Eq.~(\ref{G_lesser and G_greater}) below into Eq.~(\ref{Self_retarded}), and further approximate Eq.~(\ref{Self_retarded}) in the regime where the relaxation factor, $\eta_{1} \ll \hbar\omega$ (we take $\eta_{1}$ = 0.0033eV throughout this work); we get  
\begin{equation} \label{Self_retarded2}
\begin{aligned}
\Pi^{r}_{jj'}(\textbf{q}_{\perp},\omega)=&
\frac{2\,e^{2}}{N}\sum_{n,n'=\pm 1}\sum_{\textbf{k}}\:\Xi^{j,\,j'}_{n,\,n'}\\
&\biggl[\frac{n_{F}(\epsilon_{n'}(\textbf{k}-\textbf{q}_{\perp}))-n_{F}(\epsilon_{n}(\textbf{k}))}{\hbar\omega+\epsilon_{n'}(\textbf{k}-\textbf{q}_{\perp})-\epsilon_{n}(\textbf{k})+i\,\eta_{1}}\biggr],
\end{aligned}
\end{equation} 
where
\begin{equation}
\begin{aligned}
&\epsilon_{n}(k) = n\,\gamma_{0}|f(k)|,\\
&\Xi^{j,\,j'}_{n,\,n'}= [S_{1}]_{j\,n}[S_{2}]_{j'\,n'}[S_{1}]_{j'\,n}^{*}[S_{2}]_{j\,n'}^{*},\\
&S_{1}= \frac{1}{\sqrt{2}}
\biggl(\begin{matrix}
1 & e^{i\,\varphi(\textbf{k})}\\
-e^{-i\,\varphi(\textbf{k})} & 1
\end{matrix}\biggr), \\
&S_{2}= \frac{1}{\sqrt{2}}
\biggl(\begin{matrix}
1 & e^{i\,\varphi(\textbf{k}-\textbf{q}_{\perp})}\\
-e^{-i\,\varphi(\textbf{k}-\textbf{q}_{\perp})} & 1
\end{matrix}\biggr), \\
&\varphi(\textbf{k})= -i\ln(f(\textbf{k})/\abs{f(\textbf{k})}).
\end{aligned}
\end{equation}

We evaluate Eq.~(\ref{Self_retarded2}) numerically, instead of using the long-wave approximation formula in Refs. [\citenum{Wunsch_2006}, \citenum{Hwang_2007}].

\subsection{LEA}
Consider only one of the two sheets, the bare retarded Green's function of electrons in graphene reads 
\begin{equation}\label{G_retarded}
G^{r}(\textbf{k}, E)={}
\biggl(
\begin{matrix}
E+i\,\eta_{1}& \gamma_{0}f(\textbf{k})\\
\gamma_{0}f(\textbf{k})^{*} & E+i\,\eta_{1}
\end{matrix}
\biggl)^{-1}.
\end{equation}
In LEA, it can be assumed from fluctuation-dissipation theorem \cite{PhysRevE.75.061128,Wang2014} that the lesser and greater Green's function of electron is related to the retarded and advanced by
\begin{equation}\label{G_lesser and G_greater}
\begin{aligned}
G^{<}&(\textbf{k}, E)={} -n_{F}(E)\left(G^{r}(\textbf{k}, E)-G^{a}(\textbf{k}, E)\right),\\
G^{>}&(\textbf{k}, E)={} \left(1-n_{F}\right)\left(G^{r}(\textbf{k}, E)-G^{a}(\textbf{k}, E)\right),
\end{aligned}
\end{equation}
where $n_{F}(E)={} 1/(e^{\beta(E-\mu)}+1)$, the Fermi distribution.
Also, the lesser and greater self-energy read
\begin{equation}\label{Pi_lesser and Pi_greater}
\begin{aligned}
\Pi^{<}&(\textbf{q}_{\perp}, E)={} n_{B}(E)\left(\Pi^{r}(\textbf{q}_{\perp}, E)-\Pi^{a}(\textbf{q}_{\perp}, E)\right),\\
\Pi^{>}&(\textbf{q}_{\perp}, E)={} \left(1+n_{B}\right)\left(\Pi^{r}(\textbf{q}_{\perp}, E)-\Pi^{a}(\textbf{q}_{\perp}, E)\right),
\end{aligned}
\end{equation}
where $n_{B}(E)={} 1/(e^{\beta E}-1)$, the Bose distribution.

\section{Analytic proof of the Caroli formula in long-wave limit}\label{Analytic Proof of the Caroli Formula in Long Wave Limit}
In the long-wave limit, the A and B sublattices are indistinguishable. The transition into such limit is made by the replacement: $D_{0}^{r} \rightarrow \frac{i\,e^{i\,q_{z}\,|z-z'|}}{2\,\epsilon_{0}\,S_{c}\,q_{z}}$ and $\Pi^{r} \rightarrow \frac{1}{4} \sum_{j,\,j'}\Pi^{r}_{jj'}$. 
Equation~(\ref{the proposed current flux}) becomes
\begin{equation}\label{the proposed current flux (long-wave limit)}
\langle \hat{j}_z\rangle(z)=\, \epsilon_{0}\frac{1}{N}\sum_{ \textbf{q}_{\perp}} \int_{0}^{\infty}\frac{d\omega}{\pi}\,\hbar\omega\,
{\rm Re}\,\frac{\partial D^{>}(\textbf{q}_{\perp},\omega,z,z')}{\partial \, z'}\biggr|_{z'=z}.
\end{equation}
And $\hat{O}$ in Eq.~(\ref{Spectral Transmission}) has become $2\times2$ plate-indexed matrices this time. The trace therein is taken over plate index.

With the trace in Eq.~(\ref{Spectral Transmission}) taken explicitly, it can be further written as
\begin{equation}\label{Caroli transmission-ver 2}
T(\omega)=\int\frac{d^{2} \textbf{q}_{\perp}}{(2\pi)^2}\,
\left\{D_{21}^{r}\Gamma_{1}D_{12}^{a}\Gamma_{2}\right\},
\end{equation}
where $D_{21}^{r}=D^{r}(d,0)$ (let plate 1 locate at $z=0$ and plate 2 at $z=d$.) and so on for similar cases.

Our main aim for the comparison is just to compare
\begin{equation}\label{the part to match}
2\epsilon_{0}S_{c}\Re\,\frac{\partial}{\partial \, z'} 
D^{>}(\textbf{q}_{\perp},\omega,z,z')|_{z' =z = d^{-}},
\end{equation}
and
\begin{equation}\label{the one derived from the Caroli formula}
\left(N_{1}-N_{2}\right)\left\{D_{21}^{r}\Gamma_{1}D_{12}^{a}\Gamma_{2}\right\}.
\end{equation}

After some algebraic work, the one derived from the Caroli formula [Eq.~(\ref{the one derived from the Caroli formula})] ultimately reads
\begin{equation}\label{the part to be matched}
\left(N_{1}-N_{2}\right)\Big|\frac{D_{0\,11}^{r}}{\mathcal{L}}\Big|^{2}\,\Gamma_{1}\Gamma_{2}
\,e^{-2|q_{z}|d},
\end{equation}
with
\begin{equation*}
D_{0\,11}^{r}={}\frac{i}{2\epsilon_{0}S_{c}q_{z}}=\frac{1}{2\epsilon_{0}S_{c}|q_{z}|},
\end{equation*}
and
\begin{equation*}
\mathcal{L}= 1-\left[\tilde{\Pi}^{r}_{1}+\tilde{\Pi}^{r}_{2}
-\tilde{\Pi}^{r}_{1}\tilde{\Pi}^{r}_{2}\left(1-e^{-2\,|q_{z}|d}\right)\right].
\end{equation*}

Here We introduce a shorthand notation with $\tilde{\Pi}^{r}_{1}=D_{0\,11}^{r}\Pi^{r}_{1}$ and so forth for the like.
\begin{equation*}
\begin{aligned}
&2\epsilon_{0}S_{c} {\rm Re}\,\frac{\partial}{\partial \, z'} 
D^{>}(z,z')|_{z'=z}\\
&={\rm Re}\,\sum_{l=1,\,2}\epsilon_{0} D^{r}(z,z^{(l)}) \Pi_{l}^{>}\frac{\partial}{\partial \, z'}D^{a}(z^{(l)},z')|_{z'=z}\\
&={\rm Re}\,\sum_{l=1,\,2}\epsilon_{0} D^{r}(z,z^{(l)}) \Pi_{l}^{>}\left[\frac{\partial}{\partial\, z'}D^{r}(z',z^{(l)})|_{z'=z}\right]^{*}. 
\end{aligned}
\end{equation*} 

Taking $z\rightarrow d^{-}$,
\begin{equation}
\begin{aligned}
D^{r}(d^{-},z^{(l)})=&
\begin{cases*}
\frac{D_{0\,11}^{r}}{\mathcal{L}}e^{-|q_{z}|d}&,\, \text{$l$=1;}\\ 
\frac{D_{0\,11}^{r}}{\mathcal{L}}[1-\tilde{\Pi}^{r}_{1}\left(1-e^{-2|q_{z}|d}\right)]&,\, \text{$l$=2;} 
\end{cases*}\\
   \Pi_{l}^{>}=&
    \begin{cases*}
    -i\,\left(1+N_{1}\right)\Gamma_{1} &,\, \text{$l$=1;}\\ 
    -i\,\left(1+N_{2}\right)\Gamma_{2} &,\, \text{$l$=2;} 
    \end{cases*}\\
\frac{\partial}{\partial \, z'}D^{r}(z',z^{(l)})&|_{z'=d^{-}}\\
=&  \begin{cases*}
   -|q_{z}|\frac{D_{0\,11}^{r}}{\mathcal{L}}e^{-|q_{z}|d}\left[1-2\tilde{\Pi}^{r}_{2}\right]&,\, \text{$l$=1;}\\ 
   -|q_{z}|\frac{D_{0\,11}^{r}}{\mathcal{L}}[-1+\tilde{\Pi}^{r}_{1}\left(1+e^{-2|q_{z}|d}\right)]&,\, \text{$l$=2.} 
   \end{cases*}
\end{aligned}
\end{equation}

Putting it all together, and after taking the real part, one finally gets
\begin{equation}\label{the final one}
\begin{aligned}
&-2\epsilon_{0}S_{c}|q_{z}|\Biggl|\frac{D_{0\,11}^{r}}{\mathcal{L}}\Biggr|^{2}\times\\
&\biggl[-e^{-2|q_{z}|d}\left(1+N_{1}\right)\Gamma_{1}\left[D_{0\,11}^{r}\Gamma_{2}\right]\\
&+\left(1+N_{2}\right)\Gamma_{2}\biggl[D_{0\,11}^{r}{\rm Im}\Pi^{r}_{1}\left(1-e^{-2|q_{z}|d}\right)\\
&-D_{0\,11}^{r}{\rm Im}\Pi^{r}_{1}\left(1+e^{-2|q_{z}|d}\right)\biggr]\biggr]\\
&=\Biggl|\frac{D_{0\,11}^{r}}{\mathcal{L}}\Biggr|^{2}e^{-2|q_{z}|d}\,
\Gamma_{1}\Gamma_{2}\left(N_{1}-N_{2}\right).
\end{aligned}
\end{equation} 
There is no doubt that Eq.~(\ref{the part to match}) matches Eq.~(\ref{the part to be matched}). This shows the ``Caroli" formula is just our ``Poynting scalar" formula (Eq.~(\ref{the proposed current flux (long-wave limit)}) and Ref. [\citenum{Peng_2-dot}]).

\bibliography{manuscript}

\end{document}